\documentclass{aa}	

\usepackage{graphicx}
\usepackage[varg]{txfonts}

\usepackage{amssymb}

\usepackage{natbib}
\bibpunct{(}{)}{;}{a}{}{,}

\usepackage{ulem}

\begin{document}
\title{Solar wind He pickup ions as source of tens-of-keV/n neutral\\ He atoms observed by the HSTOF/SOHO detector} 

\author{
S.~Grzedzielski 	\inst{\ref{src}}
\and  
P.~Swaczyna 		\inst{\ref{src}}
\and  
A.~Czechowski 		\inst{\ref{src}}
\and  
M.~Hilchenbach 		\inst{\ref{mps}}
}

\institute{
Space Research Centre, Polish Academy of Sciences, Bartycka 18A, 00-716 Warsaw, Poland \\
\email{stangrze@cbk.waw.pl, pswaczyna@cbk.waw.pl, ace@cbk.waw.pl} \label{src}
\and 
Max-Planck-Institut f\"{u}r Sonnensystemforschung, Max-Planck-Strasse 2, 37191 Katlenburg-Lindau, Germany \\
\email{hilchenbach@mps.mpg.de} \label{mps}
}

\date{Received [date] / Accepted [date]}

\abstract
{Since 1996, during periods of low solar activity, the HSTOF instrument onboard the SOHO satellite has been measuring weak fluxes of He atoms of 28--58 keV/n (helium energetic neutral atoms -- He ENA). The probable source region is the inner heliosheath.}
{We aim to understand the emission mechanism of He ENA based on knowledge of heliosheath spatial extent and plasma content resulting from Voyager 1 \& 2 measurements in the period after termination shock crossings.}
{He ENA are generated by charge-exchange neutralization of energetic helium ions on interstellar neutral H and He. Energy spectra of helium ions in the heliosheath are calculated  by following the evolution of their velocity distribution functions when carried by and undergoing binary interactions with plasma constituents of a background flow whose particle populations are modeled to approximately render post-termination-shock Voyager data.}
{The observed HSTOF He ENA form a higher energy part of general heliospheric He ENA fluxes and can be explained by the proposed mechanism to within 2$\sigma$ error. The main factor determining the level of emission (and its uncertainty) is the energy spectrum of He$^+$ pickup ions in post-termination shock plasmas.}
{}


\titlerunning{Solar wind He PUIs as source of He ENA observed by the HSTOF/SOHO detector}

\authorrunning{Grzedzielski et al.}

\maketitle

\section{Introduction}
\label{introduction}

Study of energetic neutral atoms of hydrogen (H ENA) and helium (He ENA) that results from ion-neutral atom charge exchange at the confines of the heliosphere provides access to plasma conditions in the inner and outer heliosheaths. For almost two decades, very low fluxes of 58--88 keV H ENA and 28--58 keV/n He ENA of probably heliosheath origin are being measured by the HSTOF experiment onboard SOHO \citep{hilchenbach_etal:12a}. Data on these neutral particles can only be collected during quiet solar times, i.e. mainly around solar minima when the ion fluxes impinging on spacecraft are low \citep{hilchenbach_etal:98a,hilchenbach_etal:01a}. Available data cover all periods from 1996, except for a gap when SOHO was beyond reach. The data are dominated by the years 1996--1997 and the recent period of deep solar minimum (2006--2010).

The viability of using the H ENA observed by HSTOF as a means to study the heliosheath was first pointed out by \citet{hilchenbach_etal:98a}. This idea was then extended to He ENA by \citet{czechowski_etal:01a} and further developed in a series of papers \citep{czechowski_etal:06a,czechowski_etal:08a}. Because of low fluxes in the HSTOF energy range, a meaningful analysis is only possible during `quiet times' and after integrating the signal over large swaths of the sky. In the recent paper by \citet{czechowski_etal:12a}, four sectors of ecliptic longitudes were analyzed: 120$^\circ$--210$^\circ$ and 300$^\circ$--30$^\circ$ (flanks), 210$^\circ$--300$^\circ$ (apex), and 30$^\circ$--120$^\circ$ (heliotail), with latitudes confined to the $\pm17^\circ$ interval defined by the HSTOF field of view. 
The sector-integrated H and He ENA fluxes were interpreted in terms of a heliosheath model that includes effects of energetic ion convection, charge-exchange loss, longitudinal (in the sense of magnetic field) diffusion and (perpendicular) diffusive escape beyond the heliopause. However, the model of \citet{czechowski_etal:12a} predicts fluxes that are by a factor of a few higher than observed, both for H ENA and He ENA. Nevertheless, in the lowest HSTOF energy range (28--38 keV/n), where the uncertainty, in particular due to ion contamination, is at its minimum, the simulated results for He ENA were higher than observed in the heliotail, flank sector, and forward sector by only factors 1.5, 2, and 3, respectively.
In the present paper we analyze the situation again with the modeling and show that, based on heliosheath size and plasma content as emerging from the most recent Voyager 1 (V1) and Voyager 2 (V2) data, it seems possible (1) to reduce the hiatus between HSTOF He ENA observations and modeling, and (2) to perceive the HSTOF energy range as a high-energy tail of a general population of heliospheric ENA (both H and He) that now becomes accessible owing to the Insterstellar Boundary Explorer (IBEX) measurements \citep{mccomas_etal:09a}.

The possible cause of the above-mentioned discrepancy with the HSTOF He ENA data could be that the \citet{czechowski_etal:12a} calculations were based on a gas-dynamical model of the heliosphere by \citet{fahr_etal:00a}, complemented with heliospheric magnetic field and interstellar neutral He background. For obvious reasons, the \citet{fahr_etal:00a} model could take into account neither the heliosheath plasma conditions observed since 2007 by the plasma experiment onboard V2 \citep[i.e., pressure residing mainly in non-thermal ion populations, cf.][]{richardson_etal:08a} nor the reduced -- compared to earlier estimates -- distance to the heliopause along the V1 trajectory. The new value, probably $\sim122~\mathrm{AU}$, was recently inferred from the dramatic decline of heliospheric energetic particle populations observed at that distance by V1 \citep{stone_etal:13a,burlaga_etal:13a,krimigis_etal:13a,webber_mcdonald:13a}. 
Although the identification of the observed boundary as the heliopause is still in some doubt, it seems certainly to be the outer limit of the heliospheric energetic ions, so it is the bound observable in ENA.

The study of heliospheric ENA has been substantially invigorated since the December 2008 launch of the IBEX. This experiment provides all sky coverage of H ENA fluxes in a number of energy channels in the range 0.2--6 keV (6 and 8 channels in the Hi- and Lo- detectors, respectively). This development resulted in novel insight into the condition of plasma and magnetic fields at the confines of the heliosphere, both in the inner and in the outer heliosheath \citep{mccomas_etal:11a}. In particular it led to the discovery of the H ENA Ribbon that presents the most striking feature in H ENA sky observed by IBEX. Though no identification of He ENA fluxes in the IBEX energy range has as yet been reported, simple theoretical modelings of expected heliosheath He ENA fluxes were developed in view of the plausible future detection and importance of such helium data \citep{grzedzielski_etal:13a}. 
The method employed followed the general approach worked out earlier \citep{grzedzielski_etal:10a} for predicting energetic atom fluxes of heavy species in the heliosheath. Out of three simple models developed in the 2013 paper, two corresponded to the above-mentioned reduced distance to the heliopause and took account of a heliosheath filled with thermal and non-thermal plasma populations that conform with Voyager 2 post-TS (termination shock) observations \citep{richardson_wang:11a}. The models predicted that He ENA fluxes integrated over the IBEX energy range should approximately vary from $\sim0.01~\mathrm{(cm^2~s~sr~keV)^{-1}}$ (apex direction) to $\sim2~\mathrm{(cm^2~s~sr~keV)^{-1}}$ (tail direction). 
Though this seems to be somewhat below the estimated current sensitivity of IBEX instrumentation for detecting He ENA \citep{allegrini_etal:08a}, it nevertheless suggests that the level of 0.5--6 keV He ENA fluxes may not be far from the threshold of detectability. One can thus expect that He ENA measurements in the energy range of few keV/n could, in future, provide a valuable additional diagnostic tool of heliosheath plasmas. In this context it becomes important to check whether the developed modeling is also able to adequately explain the already observed He ENA fluxes in the range of energies covered by the HSTOF experiment.

To this purpose we work out an extension into the HSTOF He energy range (28--58 keV/n) of the modeling employed in \citet{grzedzielski_etal:13a} for the IBEX energy range. We show that, based on very simple modeling, the observed HSTOF He ENA fluxes can be reasonably understood as the higher energy part of the He ENA fluxes expected from a reduced heliosheath filled with plasma of the type identified by V2. In this way present results concur with the picture of a relatively small heliosphere, anticipated on the basis of H ENA data by \citet{hsieh_etal:10a} and very strongly suggested by Voyager 1 in situ observations of energetic ions. They also give support to the view that measurements of the thermal plasma population along the relatively short stretch of the V2 trajectory may indeed represent the typical conditions prevailing in heliosheath plasmas.

\section{Physical model of processes leading to emergence of tens-of-keV/n He ENA fluxes}
\label{model}

Following \citet{grzedzielski_etal:13a} we treat He ions ($\alpha$-particles and He$^+$ ions) as test particles carried hydrodynamically by a time-independent, axisymmetric bulk heliosheath plasma flow (Sect.~\ref{plasma}). The particles are described in terms of their corresponding velocity distribution functions, $f^\alpha$ and $f^\mathrm{He^+}$, which are assumed to be isotropic in velocity space in the local plasma frame. The assumption of isotropy has recently been tested and confirmed in the case of 53--85 keV ions (presumably protons) in a series of reorientations of the V1 spacecraft performed at heliocentric distances of $\sim$ 117--120 AU \citep{decker_etal:12a}. The (scalar) velocity space is divided into 500 velocity bins of $20~\mathrm{km~s^{-1}}$ width covering the range from $0~\mathrm{km~s^{-1}}$ to $1000~\mathrm{km~s^{-1}}$. Initial conditions for $f^\alpha$ and $f^\mathrm{He^+}$ are specified in the post-TS regions and are discussed in Sects.~\ref{plasma}~\&~\ref{results}. 
The particles are changing their charge states because of various binary interactions they undergo with each other and with other plasma constituents and neutral atoms present in the heliosheath.

The binary interactions (BI) we consider include radiative and dielectronic recombinations, electron impact ionizations, photoionizations, double and single charge exchanges (also to upper levels), and electron stripping. They are essentially those that were used in \citet{grzedzielski_etal:13a} (cf. Fig. 1 and Table 1 therein for list of interactions and cross section references). Two new binary interactions important for the high energy range now discussed were added, namely ionization of He$^+$ upon collision with neutral atoms of H and He. These are now the main  conversion channels of He$^+$ into $\alpha$-particles for energies $\gtrsim15~\mathrm{keV/n}$. The corresponding cross sections were taken from \citet{redbooks}. Erroneous mishandling of some transitions to excited energy states by \citet{grzedzielski_etal:13a} was also eliminated, which had no practical influence on the previously obtained results in the IBEX energy range. 

The changes in $f^\alpha$ and $f^\mathrm{He^+}$ along each of the flow lines (described by running coordinate $s$) of bulk plasma flowing with velocity $v_\mathrm{sw}(s)$ are determined by appropriate coupled transport equations for a cosmic ray type gas \citep{jokipii:87a}. We use equations of generally the same type as Eqs. (2) and (3) in \citet{grzedzielski_etal:13a}. For instance, the equation for He$^+$, which as it turns out, will now be the main ionic population effectively determining the fluxes of He ENA for the HSTOF energies, reads as
\begin{eqnarray} \nonumber
v_{\mathrm{sw}}\frac{\mathrm{d}}{\mathrm{d} s}f^{\mathrm{He}^+}&=&
G_{\mathrm{BI,}\alpha\rightarrow\mathrm{He}^+}-L_{\mathrm{BI,He}^+\rightarrow\alpha}-L_{\mathrm{BI,He}^+\rightarrow\mathrm{He}}\\
&&-L_{\mathrm{C,}\nu_{\epsilon}^{\mathrm{He}^+ \backslash\mathrm{p}}}-L_{\mathrm{He}^+,\mathrm{H}}
-L_\mathrm{esc,He^+}\, .
\label{fheplus}
\end{eqnarray}

The successive terms on the righthand-side of Eq.~\eqref{fheplus} describe changes in $f^\mathrm{He^+}$ due to gain ($G_\mathrm{BI,\alpha\rightarrow He^+}$) from BI conversion of $\alpha$ into He$^+$, loss ($L_\mathrm{BI,He^+\rightarrow\alpha}$) from BI conversion of He$^+$ into $\alpha$, loss ($L_\mathrm{BI,He^+\rightarrow He}$) from BI conversion of He$^+$ into He, loss ($L_{\mathrm{C,}\nu_{\epsilon}^{\mathrm{He}^+ \backslash\mathrm{p}}}$) due to Coulomb scattering on background protons corresponding to energy loss rate $\nu_{\epsilon}^{\mathrm{He}^+ \backslash\mathrm{p}}$ as given by \citet{huba:02a}, and loss ($L_\mathrm{He^+,H}$) due to He$^+$ interaction with neutral hydrogen \citep{berger_etal:05a}. These new important contributions to He$^+$ ionization by impact with neutral H and He atoms are included in the $L_\mathrm{BI,He^+\rightarrow\alpha}$ term. 
The last term, $L_\mathrm{esc,He^+}= f^\mathrm{He^+}/\tau_\mathrm{esc}$, describes diffusive escape through the heliopause with escape time $\tau_\mathrm{esc}$. It was added following \citet{czechowski_etal:12a}. Time $\tau_\mathrm{esc}$ corresponds to a random walk over the shortest distance $l_\mathrm{HP}$ separating the considered point from the heliopause. Since this diffusion has to proceed across the magnetic field, the assumed mean free path for the process is small, 0.1 AU \citep{czechowski_etal:12a}, which is consistent with the usual assumption that the transverse mean free path is on the order of $10^{-2}$ of the parallel mean free path \citep[e.g.,][]{potgieter:13a}. Analogous equation determines the evolution of $f^\alpha$
\begin{eqnarray} \nonumber
v_{\mathrm{sw}}\frac{\mathrm{d}}{\mathrm{d} s}f^{\alpha}&=&
G_{\mathrm{BI,He}^+\rightarrow\alpha}-L_{\mathrm{BI,}\alpha\rightarrow\mathrm{He}^+}-L_{\mathrm{BI,}\alpha\rightarrow\mathrm{He}}\\
&&-L_{\mathrm{C,}\nu_{\epsilon}^{\alpha \backslash\mathrm{p}}}-L_{\alpha,\mathrm{H}}
-L_\mathrm{esc,\alpha}\, .
\label{falpha}
\end{eqnarray}
   
Compared to \citet{grzedzielski_etal:13a}, we now retain neither adiabatic heating/cooling nor diffusive terms related to the density variations of the background plasma. This is the consequence of the very simple bulk flow used with constant background plasma density (which is assumed to be a fair representation of the overall thermal plasma density in the heliosheath, cf. Sect.~\ref{plasma}). Consistent with that and also following \citet{grzedzielski_etal:13a} and \citet{czechowski_etal:12a}, we do not consider Fokker-Planck type stochastic acceleration. Therefore, as long as a particle does not undergo a binary interaction, it may only undergo shifts to lower velocity bins (loss of energy $L_{\mathrm{C,}\nu_{\epsilon}^{\mathrm{He}^+\backslash\mathrm{p}}}+L_{\mathrm{He}^+,\mathrm{H}}$). 
The upward shifts may take place formally only when particles change population, for instance, when He$^+$ ions occupying a high velocity bin become $\alpha$-particles upon loss of an electron. These shifts are not very important. In the outcome there is relatively little evolution of the energy spectra of He ions when plasma parcels are carried from the TS to tail of the heliopause. It is the initial energy spectrum of the tens-of-keV/n He ions at the TS that actually turns out now to be the main factor determining the final HSTOF He ENA signal (cf. Sect.~\ref{results}).

\section{Hydrodynamical model of background plasma flow and the flux of He pick-up ions at the termination shock}
\label{plasma}

We chose a simple model description of the background plasma flow in which we assume there are no significant spatial variations in the large scale density distribution. This assumption agrees with the published plasma density data based on Voyager 2 measurements. For instance, if one excludes the region immediately beyond the termination shock, the net change in average density ($\sim0.0018~\mathrm{cm^{-3}}$) of thermal plasma density is less than 20\% over the 18 AU long stretch of V2 trajectory \citep[][and recent data\footnote{Recent Voyager 2 data from \url{ftp://space.mit.edu/pub/plasma/vgr/v2/daily/}.}]{richardson_wang:11a}. It is also indirectly supported by magnetic field data. Namely, the measurements by V1 show the constancy of the average magnetic field when averaged over sections of corresponding trajectories not shorter than a few AU \citep{burlaga_ness:12a,burlaga_etal:10a}. 
One might expect that large scale heliospheric density variations should be accompanied by corresponding changes in average magnetic field strength, but this is not what is observed. Along the trajectory of V1, the constancy of the average magnetic field continues until the spacecraft reaches the so-called `stagnation region' about 5 AU before the heliopause \citep{burlaga_etal:13a,webber_mcdonald:13a}. 

To describe background flow with constant density we employ the axisymmetric model by \citet{suess_nerney:90a}. This model is a variation of the classical analytical Parker model \citep{parker:61a}. In \citet{grzedzielski_etal:13a}, we used a set of three models of which the Parker model was the simplest. In the Parker model, the distance Sun--TS is assumed to be negligibly small compared to the distance Sun--heliopause (HP). In the Suess and Nerney version, this assumption can be relaxed and both distances, Sun--TS and Sun--HP, are considered final. The TS is, however, still spherical. As numerical values determining the effective solution for the heliospheric plasma flow, we take the values as in Table~\ref{table:1}.

\begin{table}
\caption{Numerical values defining the adopted background flow description}
\label{table:1}
\centering
\begin{tabular}{ll}
\hline
supersonic solar wind velocity		& $60~\mathrm{km~s^{-1}}$\\
solar wind flux at 1 AU			& $2\times10^8~\mathrm{protons~cm^{-2}s^{-1}}$\\
Sun--TS distance			& 83.7~AU\\
post-TS velocity			& $150~\mathrm{km~s^{-1}}$\\
HP--TS distance at V1 trajectory	& 27.5~AU\\
\hline
\end{tabular}
\end{table}

In the solution, the constant heliospheric plasma density is then $0.002~\mathrm{cm^{-3}}$. This reasonably agrees with the average value of $\sim0.0018~\mathrm{cm^{-3}}$ coming up from the V2 measurements \citep{richardson_wang:11a}.

The solution allows the total flux of singly ionized pickup ions of helium resulting from ionization of interstellar neutral helium in the supersonic solar wind to be estimated at the termination shock. We took an interstellar helium density distribution as given by \citet{rucinski_etal:98a} and assumed \citep[following][]{bzowski_etal:13a} that photoionization is the primary ionization mechanism and that the rate at 1 AU (averaged over the last solar cycle) is $9.0\times10^{-8}~\mathrm{s^{-1}}$. The average He$^+$ PUI density downstream of the TS is then $1.6\times10^{-5}~\mathrm{cm^{-3}}$ in the apex direction. The flux towards the heliospheric tail is twice as large as the flux in the apex direction. This is caused by solar gravitational focusing of the neutral interstellar He.

Once the total flux of He$^+$ ions at the TS is known, the most important factor that effectively determines the spectral density of parent ions for the HSTOF He ENA is the initial (i.e., at the TS) energy spectrum of He$^+$ ions. Then further evolution of this spectrum is determined by Eqs.~\eqref{fheplus} and~\eqref{falpha}. We identify the initial spectrum with the spectrum of ions observed in situ by V1 behind the TS. The difficulty here is that while it is obvious that spectral energy distribution is close to a power law with an exponent (index) $\gamma=-1.65$, there is an uncertainty of about a factor three concerning the absolute intensity of the spectrum of He$^+$. 

\begin{figure*}
\sidecaption
\includegraphics[width=12cm]{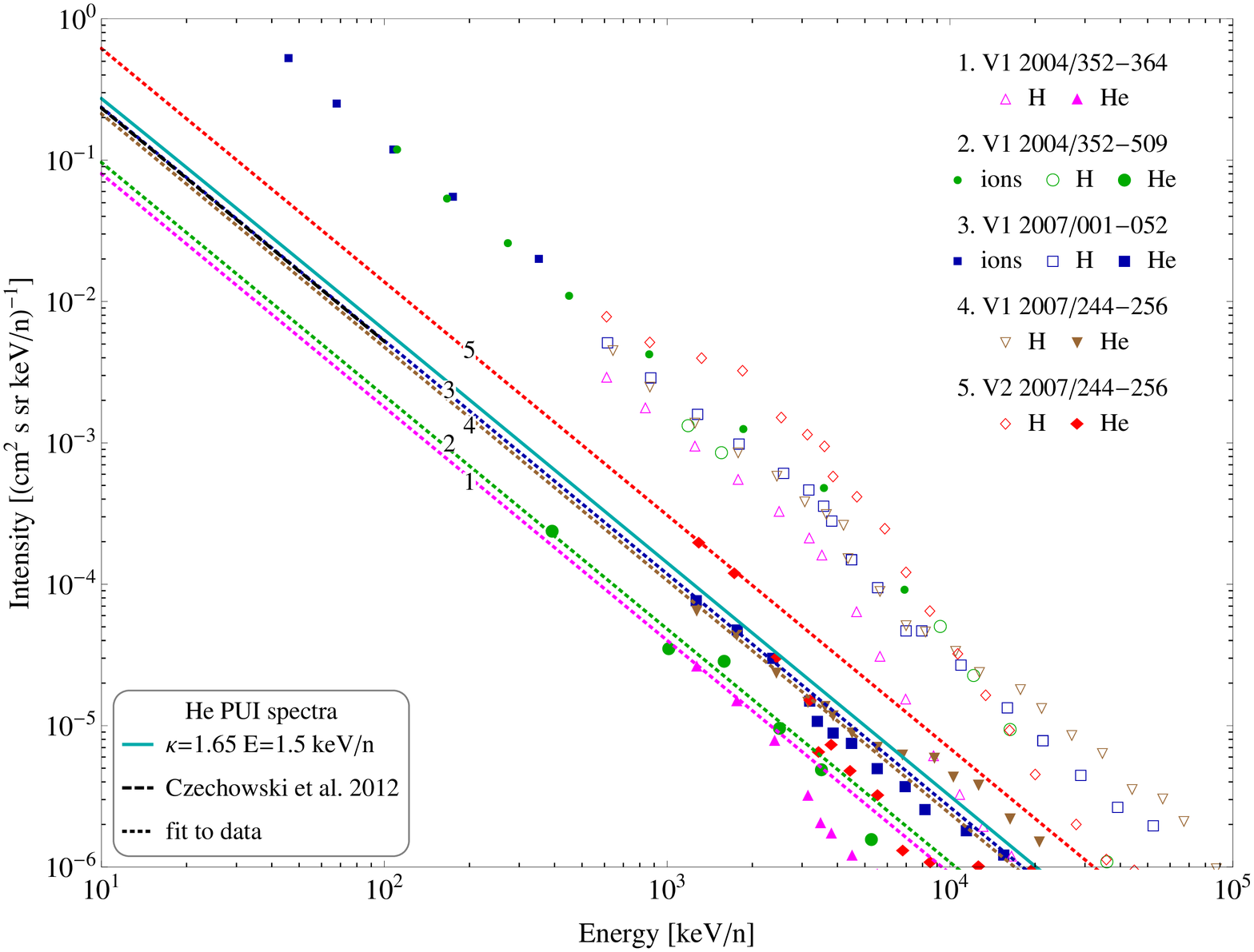}
\caption{Heliosheath He ions spectral intensities based on published data. Oblique colored (numbered) dotted lines correspond to low-energy ($<2~\mathrm{MeV/n}$, $\gamma = -1.65$) fits to He data obtained at different epochs as indicated in the upper right corner: 1. magenta (up triangle) -- V1 \citep{stone_etal:08b}; 2. green (circle) -- V1 \citep{decker_etal:05a}; 3. dark blue (square) -- V1 \citep{stone_etal:08a}; 4. brown (down triangle) -- V1 \citep{stone_etal:08b}; and 5. red (diamond) -- V2 \citep{stone_etal:08b}. He/H data are denoted by filled/empty symbols and ions data by small symbols. Solid light blue line describes a kappa distribution with $\kappa=1.65$ and average He$^+$ ion energy at the termination shock equal to 1.5 keV/n. Short black dashed line in the upper left corner shows (part of) the run of the initial He spectrum used by \citet{czechowski_etal:12a}. }
\label{hepui}
\end{figure*}

In Fig.~\ref{hepui} we show the available heliosheath ion spectral intensities based on published data by \citet{decker_etal:05a} and \citet{stone_etal:08a,stone_etal:08b}. Low-energy ($<2~\mathrm{MeV/n}$, $\gamma = -1.65$) fits to these data are shown by dotted lines of different colors (Fig.~\ref{hepui}). Also shown are a kappa distribution with $\kappa=1.65$ and average He$^+$ ion energy at the TS equal to 1.5 keV/n and the run of the initial He spectrum used in the calculations by \citet{czechowski_etal:12a} (upper left corner in Fig.~\ref{hepui}). In the present paper we performed integration of Eqs.~\eqref{fheplus} and~\eqref{falpha} for all initial spectra fitted to the Voyager data and also for the mentioned kappa distribution.        

\section{Expected He ENA intensity spectra and comparison with the HSTOF measurements in the 28--58 keV/n energy range}
\label{results}

Integration of Eqs.~\eqref{fheplus} and~\eqref{falpha}, based on described background plasma model and initial conditions at the termination shock (Sect.~\ref{plasma}), yields the distribution functions $f^\alpha$ and $f^\mathrm{He^+}$ as functions of velocity (energy per nucleon) and position in the heliosheath. In the tailward direction, the integration was performed to heliocentric distance of 15\,000 AU. To calculate the source function for the ENA fluxes, one has to know the density distribution of the interstellar neutral atoms of H and He, whose charge exchange with energetic He-ions gives rise to He ENA intensities. 

To describe the distribution of neutral interstellar H in the heliosheath we used three models: 
\begin{enumerate}[(a)]
 \item Internal part of the original model determined by Monte Carlo calculations for a large heliosphere (apex heliopause at 177 AU from the Sun) by \citet{izmodenov_alexashov:03a}; neutral H density `at infinity' $=0.2~\mathrm{cm^{-3}}$.
 \item \citet{izmodenov_alexashov:03a} model but rescaled to the presently assumed size of the heliosphere (apex heliopause at 110 AU from the Sun); neutral H density ‘at infinity’ as above.
 \item Model based on numerical retrieval of the published interstellar neutral H distribution data used by \citet{zank_etal:13a}. In our axisymmetric calculations, we used a neutral H distribution corresponding to midway density values between the ecliptic and polar cross sections of their Model 2; neutral H density at the termination shock $0.1~\mathrm{cm^{-3}}$.
\end{enumerate}
The density of neutral interstellar He in the heliosheath was assumed to be constant and equal to $0.015~\mathrm{cm^{-3}}$.

The local emissivity (source function) $j_\mathrm{ENA}$ $\mathrm{(cm^3 s~sr~keV/n)^{-1}}$ of He ENA is determined by the product of reagents' densities and reaction rates (i.e., relative velocity times cross section, $\mathrm{cm^3s^{-1}}$) for binary interactions that convert He ions into neutral He atoms by charge exchange. The most important contribution to the source function comes from He$^+$ charge exchange with neutral H with small additions from He$^+$ (single) charge exchange and the $\alpha$-particle double charge exchange with neutral He atoms. Other binary interactions are negligible \citep[for details and cross sections see][Sect.~2.3]{grzedzielski_etal:13a}.

The He ENA intensity $I_\mathrm{ENA}(E,\theta)\ \mathrm{(cm^2 s~keV/n)^{-1}}$ of particles of energy $E$ coming from the direction at angle $\theta$ from the apex is expressed as a line-of-sight integral of the source function from the termination shock to heliopause
\begin{eqnarray} \nonumber
 I_{\mathrm{ENA}}(E,\theta)&=&
 L_{\mathrm{sw}}\int_{r_{\mathrm{TS}}}^{r_{\mathrm{HP}}}
 \frac{j_{\mathrm{ENA}}(E,r,\theta)}{4\pi}
 F_{\mathrm{CG}}(E,\theta,\vec{v}_{\mathrm{sw}})\\
 &&\qquad\qquad\qquad\times\exp\left[ -\int_{r_\mathrm{TS}}^{r} \frac{\tau_{\mathrm{ext}}(r')}{v_{\mathrm{ENA}}} \mathrm{d}r'\right]
 \mathrm{d}r \, .
 \label{Iena}
\end{eqnarray}

Equation~\eqref{Iena} contains corrections for reionization losses in the heliosheath, as well as a correction ($F_\mathrm{CG}$) for the Compton-Getting effect of plasma frame movement in the observer frame
\begin{equation}
 F_\mathrm{CG}(E,\theta,\vec{v}_\mathrm{sw})=\frac{|\vec{v}_\mathrm{ENA}|(\vec{v}_\mathrm{ENA}^2-\vec{v}_\mathrm{ENA}\cdot\vec{v}_\mathrm{sw})}{|\vec{v}_\mathrm{ENA}-\vec{v}_\mathrm{sw}|^{3}}\, ,
\end{equation}
where $\vec{v}_\mathrm{ENA}(E,\theta)$ is the ENA velocity and $\vec{v}_\mathrm{sw}$ the solar wind plasma velocity in the observer frame.

The losses in the heliosheath represented by the effective depth $\tau_\mathrm{ext}$ for extinction of He ENA are mainly due to He (single) charge exchange with, and electron stripping on, H atoms and double charge exchange with $\alpha$-particles. In view of the low \citep[$<10~\mathrm{eV}$,][]{richardson_wang:11a} electron temperature in the heliosheath, the reionization by electron impact is unimportant.

\begin{figure*}
\centering$
\begin{array}{c c}
   \includegraphics[width=0.48\textwidth]{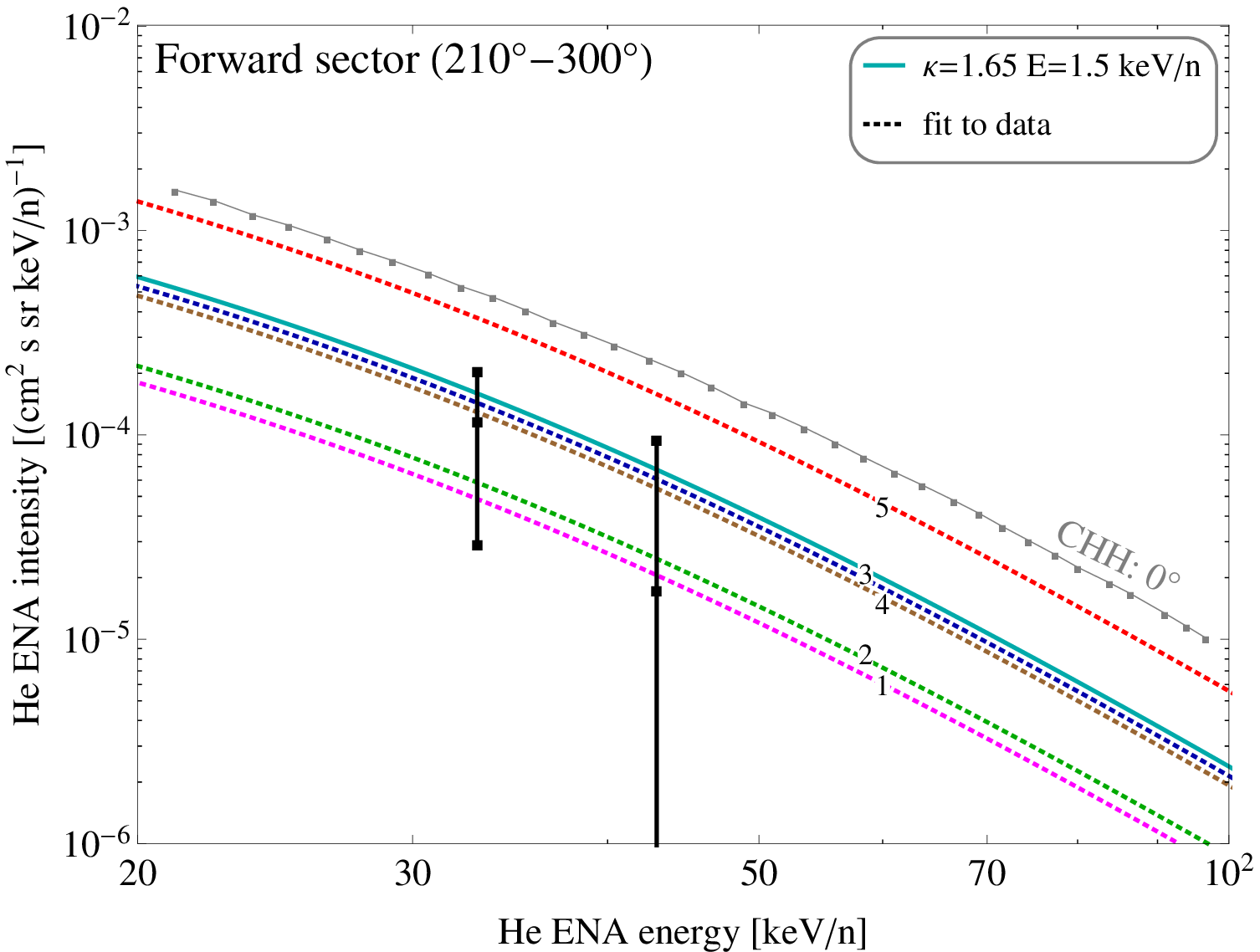} &
   \includegraphics[width=0.48\textwidth]{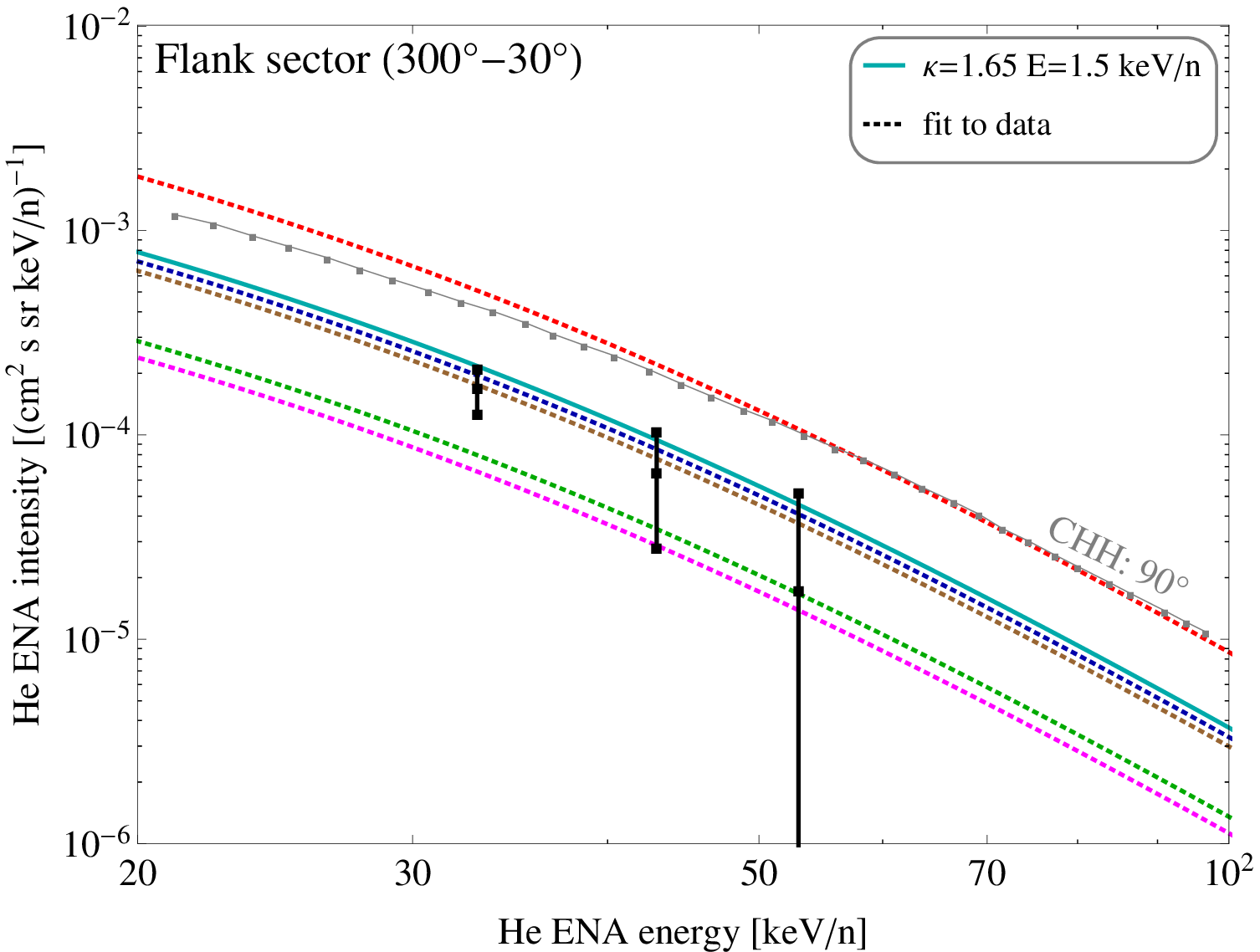} \\
   \mathrm{(a)\ Forward\ sector\ (210^\circ-300^\circ)} & \mathrm{(b)\ Flank\ sector\ (300^\circ-30^\circ)}\\
   \includegraphics[width=0.48\textwidth]{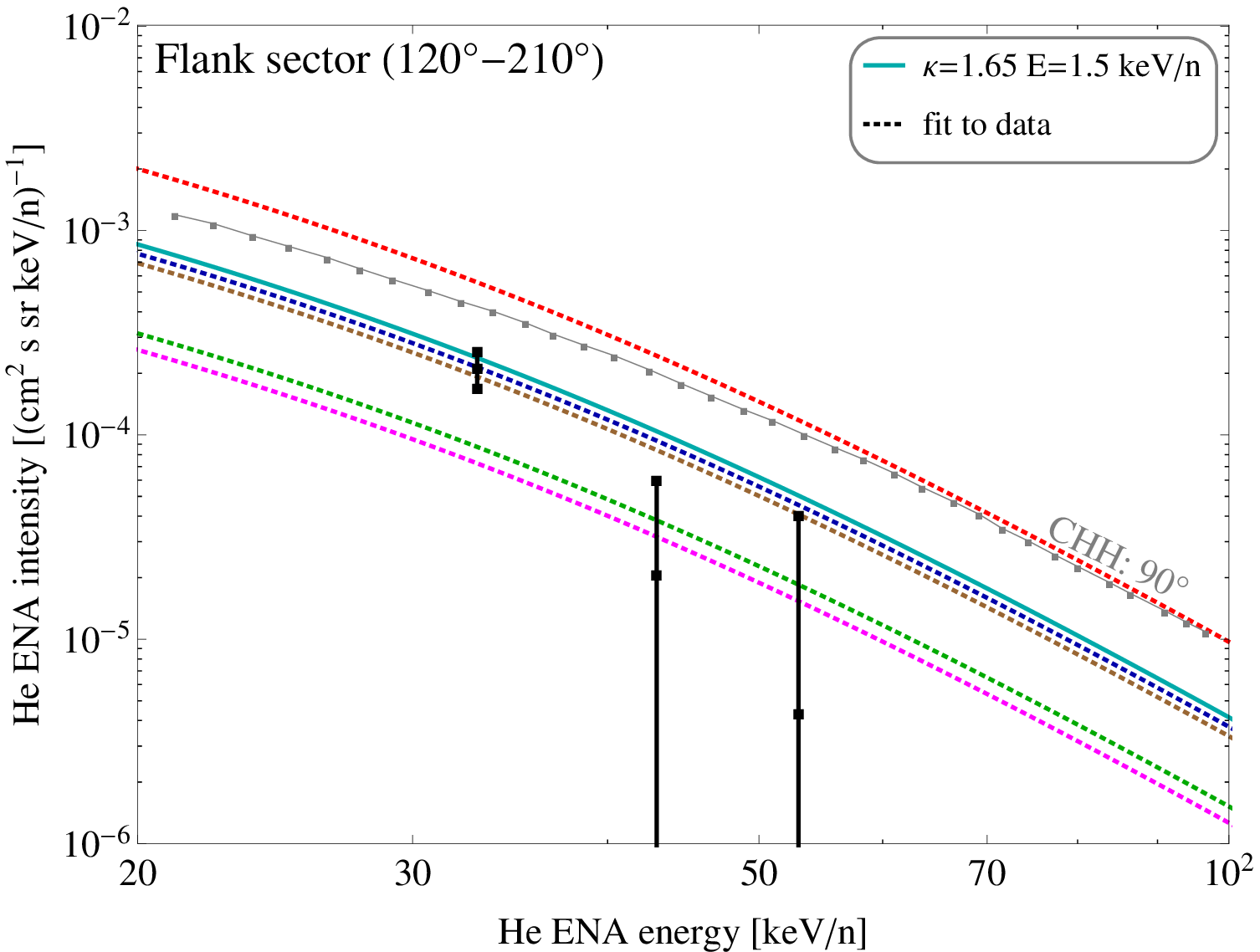} &
   \includegraphics[width=0.48\textwidth]{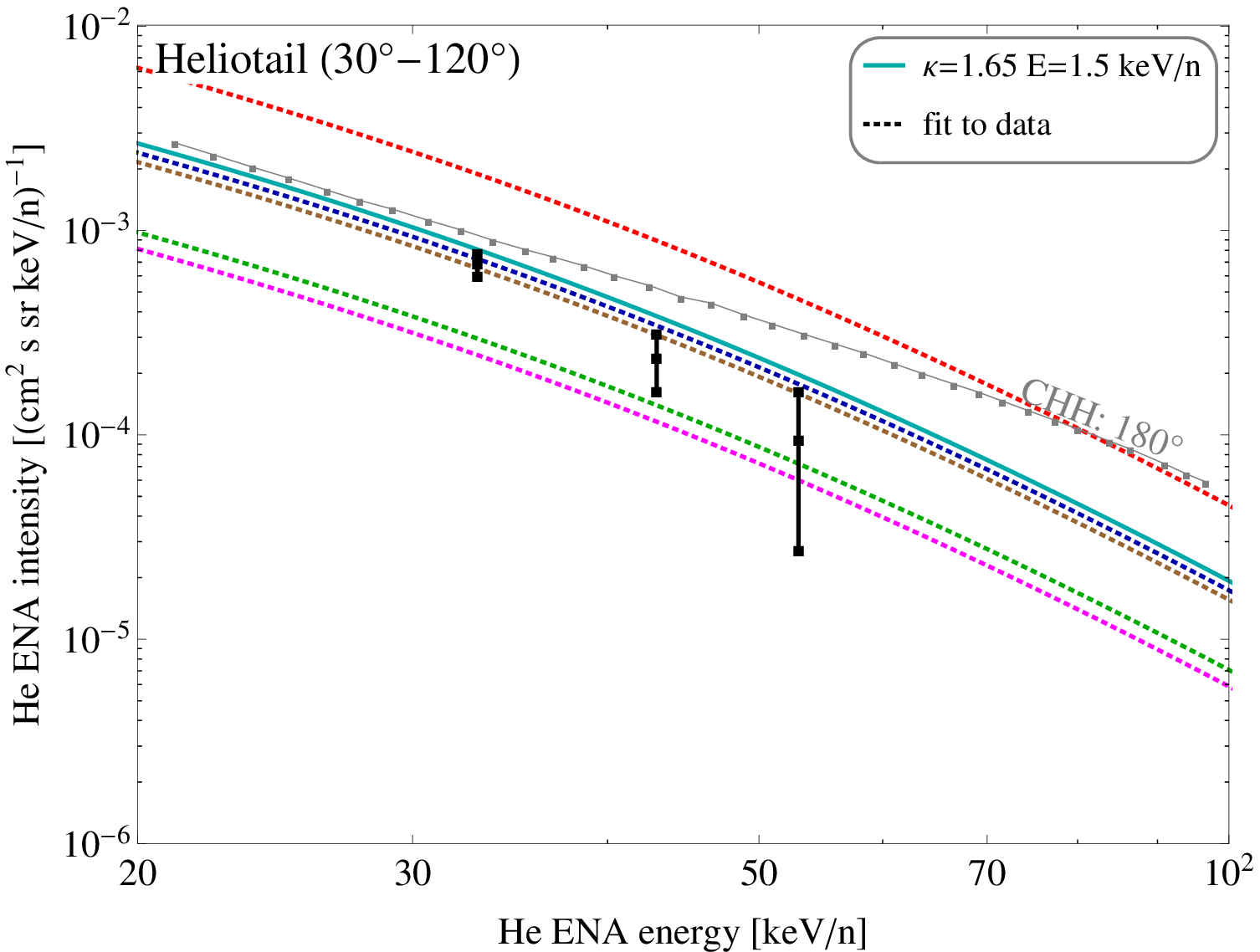} \\
   \mathrm{(c)\ Flank\ sector\ (120^\circ-210^\circ)} & \mathrm{(d)\ Heliotail\ (210^\circ-300^\circ)} \\
\end{array}$
   \caption{Comparison of He ENA spectral intensities obtained in present modeling with HSTOF He ENA measurements. The measurements were obtain in periods: forward sector 1996--2005, both flank sectors 1996--2010, tail sector 1996--2005. Dotted color curves describe calculated He ENA spectral intensities corresponding to He ion initial spectra as described by the same color in Fig.~\ref{hepui}. To aid identification, a numbering 1--5 is added to Fig.~\ref{intensity}a as in Fig.~\ref{hepui}. The solid light blue line shows the He ENA spectrum that would come out if the initial spectrum were in the form of a kappa distribution with $\kappa = 1.65$ and average He energy 1.5 keV/n. The solid gray lines with points (labeled with CHH and the value of angle from apex) correspond to previous theoretical He ENA spectra obtained by \citet{czechowski_etal:12a}.}
   \label{intensity}
\end{figure*}

To compare the expected He ENA intensities with the HSTOF data, the calculated intensities $I_\mathrm{ENA}(E,\theta)$ were averaged over four ecliptic longitude sectors as used by \citet{czechowski_etal:12a}: 210$^\circ$--300$^\circ$ (apex, Fig.~\ref{intensity}a), 120$^\circ$--210$^\circ$ and 300$^\circ$--30$^\circ$ (flanks, Fig.~\ref{intensity}b-c) and 30$^\circ$--120$^\circ$ (heliotail, Fig.~\ref{intensity}d). We took the apex orientation according to the flow direction of neutral interstellar He obtained by \citet{bzowski_etal:12a} from the IBEX measurements. Experimental HSTOF data points are shown with estimated error bars. The theoretical He ENA intensity spectra calculated with the present model are indicated as in Fig.~\ref{hepui}. The theoretical spectra are coded with numbers 1--5 (cf. Fig.~\ref{intensity}a) which correspond to numeration preceding spacecraft and date identification in Fig.~\ref{hepui}.
For comparison we also show He ENA spectra (labeled CHH) obtained previously by \citet{czechowski_etal:12a}.

It is evident that the spread of theoretical He ENA spectra presented in Fig.~\ref{intensity} is quite large, depending on the assumed initial spectra. This reflects the real spread of He ion data measured by V1 and V2. Since the HSTOF He data actually correspond to some average over various solar conditions and many years, one could presume that a safe way would be to attach more meaning to midway initial He ion values that result in He ENA spectra described by midway curves ``3'' and ``4'' in Figs.~\ref{intensity}a-d. The midway curves agree best with the observations for lowest energies for which the experimental error bars are the smallest. The hiatus between theory and observations noted by \citet{czechowski_etal:12a} is now reduced by about a factor of 3, at most, especially for the forward sector. A much better agreement -- to within 2$\sigma$ error -- between observations and present model is now evident. 
It is also interesting that the slopes of theoretical spectra in the tail direction are now somewhat closer to the observed ones than in the previous modeling. In that sense the present model probably renders the physical state of the medium better. Concerning the drawbacks, one may see that, on the whole, the slopes of the theoretical spectra of He ENA are too low (i.e., the spectra are too hard) compared with observations. However, error bars are so large that -- with the possible exclusion of the tail sector -- the slope of the observed spectrum is a data feature carrying very little weight. Overall, one may feel encouraged by the circumstance that the best agreement between present modeling and observations, both for absolute intensities and the possible slope of the spectrum, is attained in the heliotail sector for which the data uncertainties are the smallest. 

\section{Discussion and conclusions}
\label{discussion}

The main point of the present paper is to argue that based on new knowledge of the spatial extent of the heliosheath and plasma content that comes out from Voyager 1 and 2 post-TS measurements, a simple mechanism can explain the magnitude of He ENA fluxes observed by the HSTOF experiment onboard SOHO. With this mechanism the fluxes in question represent the higher energy part of a general population of heliospheric He ENA. In our model the solar wind He$^+$ pickup ions and $\alpha$-particles carried by the general hydrodynamic plasma flow in the heliosheath interact in binary collisions with each other and with the remaining background plasma constituents (Sect.~\ref{model}), to finally convert to He ENA fluxes that, after losses along the way, can be observed by HSTOF (Sect.~\ref{results}). 
The background plasma flow serving as the `playground' for these processes was described by a modification of the classical Parker model, which is chosen so as to approximately render the heliosheath size and plasma properties in concordance with the V1 and V2 measurements (Sect.~\ref{plasma}). It turns out in our modeling that the spectral intensities of generated He ENA fluxes depend crucially on the (initial) energy spectrum of He$^+$ pickup ions in post-TS plasmas, and we assume that V1 and V2 measurements of He ions spectra in the heliosheath provide a reasonable measure of these initial He$^+$ spectra. 
To compare our calculations with observations, we averaged the resulting spectral He ENA intensities over four ecliptic sectors (90$^\circ$ in longitude $\times$ 34$^\circ$ in latitude) and compared them with the HSTOF measured values in Fig.~\ref{intensity}. For comparison the result of previous modeling by \citet{czechowski_etal:12a} was also shown.

The spread of the expected spectral intensities we obtain is caused by the spread of the He ion spectra we take as the initial data for the evolution of He ion populations in the heliosheath (cf. Fig.~\ref{hepui}). We identify these initial spectra with the data measured by V1 and V2 during the few years after crossing the termination shock that is a short period compared with the $\sim10^9~\mathrm{s}$ plasma residence time in the heliosheath. The spread of these initial ion spectra is probably due to temporal changes in the heliosheath resulting from solar wind variations, possibly intertwined with spatial gradients caused by the inclination of the local interstellar magnetic field with respect to the interstellar velocity vector. 
As there is at present no clear guidance as to which of the initial He ion spectra represent the most typical case, we take midway values (Fig.~\ref{hepui}) as a possible compromise. These initial spectra generate He ENA spectral intensities described by solutions ``3'' and ``4'' in Fig.~\ref{intensity}. We feel that these solutions represent the `best approximation' to reality attainable with current modeling.

Comparison of our `best approximation' with the HSTOF data shows that improvement (in terms of intensity) over the previous model by \citet{czechowski_etal:12a}  is most significant for the forward sector (Fig.~\ref{intensity}a), less for the flank sectors (Figs.~\ref{intensity}b-c), and the least for the tail sector (Fig.~\ref{intensity}d). This improvement is mainly due to our assuming a thinner heliosheath, in agreement with recent V1 detection of the energetic ion `cliff' \citep{webber_mcdonald:13a}. One encouraging circumstance is that agreement to within $\sim20\%$ occurs in all four sectors at the lowest energies that are measured with highest accuracy. 

An additional result that could be important is that He ENA spectra obtained under current modeling are somewhat softer than the previous CHH spectra (Fig.~\ref{intensity}). This means that our spectra may perhaps render the suggested observational spectral slopes better. The softening of spectra results from including two previously neglected binary interactions in our modeling, namely (1) conversion of He$^+$ into an $\alpha$-particle upon impact of neutral H and (2) charge exchange neutralization of He$^+$ in collisions with neutral He. The effect of these two reactions should be more pronounced, the longer He$^+$ ions are exposed to binary interactions, i.e., the deeper the sources of He ENA reside in the heliotail. 
In this respect it is significant that the best agreement between the HSTOF He ENA measurements and present modeling, in terms of both intensity and slope of He ENA spectra, is attained in the heliotail sector for which, it is worth noting, the observational errors are the smallest. We take it as an indication that the type of modeling we propose in conjunction with He ENA measurements may, if properly developed, serve as diagnostic tool of the properties of heliosheath plasma at large distances from the Sun, in particular, deep in heliospheric tail.

\begin{acknowledgements} 
PS \& AC acknowledge support of the Polish National Science Centre grant 2012-06-M-ST9-00455, and the Polish Ministry for Science and Higher Education grant N-N203-513-038. 
\end{acknowledgements}

\bibliographystyle{aa} 
\bibliography{HeENAHSTOF} 

\begin{thebibliography}{38}
\expandafter\ifx\csname natexlab\endcsname\relax\def\natexlab#1{#1}\fi

\bibitem[{{Allegrini} {et~al.}(2008){Allegrini}, {Ebert}, {Alquiza}, {Broiles},
  {Dunn}, {McComas}, {Silva}, {Valek}, \& {Westlake}}]{allegrini_etal:08a}
{Allegrini}, F., {Ebert}, R.~W., {Alquiza}, J., {et~al.} 2008, Review of
  Scientific Instruments, 79, 096107

\bibitem[{{Barnett}(1990)}]{redbooks}
{Barnett}, C.~F., ed. 1990, Atomic Data for Fusion, Vol.~1 (Oak Ridge National
  Laboratory)

\bibitem[{{Berger} {et~al.}(2005){Berger}, {Coursey}, {Zucker}, \&
  {Chang}}]{berger_etal:05a}
{Berger}, M.~J., {Coursey}, J.~S., {Zucker}, M.~A., \& {Chang}, J. 2005, NIST
  Standard Reference Database 124: Stopping-Power and Range Tables for
  Electrons, Protons, and Helium Ions, {National Institute of Standards and
  Technology, Physical Measurement Laboratory}

\bibitem[{{Burlaga} \& {Ness}(2012)}]{burlaga_ness:12a}
{Burlaga}, L.~F. \& {Ness}, N.~F. 2012, \apj, 749, 13

\bibitem[{{Burlaga} {et~al.}(2013){Burlaga}, {Ness}, \&
  {Stone}}]{burlaga_etal:13a}
{Burlaga}, L.~F., {Ness}, N.~F., \& {Stone}, E.~C. 2013, Science, 341, 147

\bibitem[{{Burlaga} {et~al.}(2010){Burlaga}, {Ness}, {Wang}, {Sheeley}, \&
  {Richardson}}]{burlaga_etal:10a}
{Burlaga}, L.~F., {Ness}, N.~F., {Wang}, Y.-M., {Sheeley}, N.~R., \&
  {Richardson}, J.~D. 2010, Journal of Geophysical Research (Space Physics),
  115, 8107

\bibitem[{{Bzowski} {et~al.}(2012){Bzowski}, {Kubiak}, {M{\"o}bius},
  {Bochsler}, {Leonard}, {Heirtzler}, {Kucharek}, {Sok{\'o}{\l}}, {H{\l}ond},
  {Crew}, {Schwadron}, {Fuselier}, \& {McComas}}]{bzowski_etal:12a}
{Bzowski}, M., {Kubiak}, M.~A., {M{\"o}bius}, E., {et~al.} 2012, \apjs, 198, 12

\bibitem[{{Bzowski} {et~al.}(2013){Bzowski}, {Sok{\'o}{\l}}, {Kubiak}, \&
  {Kucharek}}]{bzowski_etal:13a}
{Bzowski}, M., {Sok{\'o}{\l}}, J.~M., {Kubiak}, M.~A., \& {Kucharek}, H. 2013,
  \aap, 557, A50

\bibitem[{{Czechowski} {et~al.}(2001){Czechowski}, {Fichtner}, {Grzedzielski},
  {Hilchenbach}, {Hsieh}, {Jokipii}, {Kausch}, {Kota}, \&
  {Shaw}}]{czechowski_etal:01a}
{Czechowski}, A., {Fichtner}, H., {Grzedzielski}, S., {et~al.} 2001, \aap, 368,
  622

\bibitem[{{Czechowski} {et~al.}(2012){Czechowski}, {Hilchenbach}, \&
  {Hsieh}}]{czechowski_etal:12a}
{Czechowski}, A., {Hilchenbach}, M., \& {Hsieh}, K.~C. 2012, \aap, 541, A14

\bibitem[{{Czechowski} {et~al.}(2008){Czechowski}, {Hilchenbach}, {Hsieh},
  {Grzedzielski}, \& {K{\'o}ta}}]{czechowski_etal:08a}
{Czechowski}, A., {Hilchenbach}, M., {Hsieh}, K.~C., {Grzedzielski}, S., \&
  {K{\'o}ta}, J. 2008, \aap, 487, 329

\bibitem[{{Czechowski} {et~al.}(2006){Czechowski}, {Hilchenbach}, \&
  {Kallenbach}}]{czechowski_etal:06a}
{Czechowski}, A., {Hilchenbach}, M., \& {Kallenbach}, R. 2006, in The Physics
  of the Heliospheric Boundaries, ed. V.~V. {Izmodenov} \& R.~{Kallenbach}, 311

\bibitem[{Decker {et~al.}(2012)Decker, Krimigis, Roelof, \&
  Hill}]{decker_etal:12a}
Decker, R.~B., Krimigis, S.~M., Roelof, E.~C., \& Hill, M.~E. 2012, \nat, 489,
  124

\bibitem[{{Decker} {et~al.}(2005){Decker}, {Krimigis}, {Roelof}, {Hill},
  {Armstrong}, {Gloeckler}, {Hamilton}, \& {Lanzerotti}}]{decker_etal:05a}
{Decker}, R.~B., {Krimigis}, S.~M., {Roelof}, E.~C., {et~al.} 2005, Science,
  309, 2020

\bibitem[{{Fahr} {et~al.}(2000){Fahr}, {Kausch}, \& {Scherer}}]{fahr_etal:00a}
{Fahr}, H.~J., {Kausch}, T., \& {Scherer}, H. 2000, \aap, 357, 268

\bibitem[{{Grzedzielski} {et~al.}(2013){Grzedzielski}, {Swaczyna}, \&
  {Bzowski}}]{grzedzielski_etal:13a}
{Grzedzielski}, S., {Swaczyna}, P., \& {Bzowski}, M. 2013, \aap, 549, A76

\bibitem[{{Grzedzielski} {et~al.}(2010){Grzedzielski}, {Wachowicz}, {Bzowski},
  \& {Izmodenov}}]{grzedzielski_etal:10a}
{Grzedzielski}, S., {Wachowicz}, M.~E., {Bzowski}, M., \& {Izmodenov}, V. 2010,
  \aap, 512, A72

\bibitem[{{Hilchenbach} {et~al.}(2001){Hilchenbach}, {Hsieh}, {Hovestadt},
  {Kallenbach}, {Czechowski}, {M{\"o}bius}, \&
  {Bochsler}}]{hilchenbach_etal:01a}
{Hilchenbach}, M., {Hsieh}, K.~C., {Hovestadt}, D., {et~al.} 2001, in The Outer
  Heliosphere: The Next Frontiers, ed. K.~{Scherer}, H.~{Fichtner}, H.~J.
  {Fahr}, \& E.~{Marsch}, 273

\bibitem[{{Hilchenbach} {et~al.}(1998){Hilchenbach}, {Hsieh}, {Hovestadt},
  {Klecker}, {Gruenwaldt}, {Bochsler}, {Ipavich}, {Buergi}, {Moebius}, {Gliem},
  {Axford}, {Balsiger}, {Bornemann}, {Coplan}, {Galvin}, {Geiss}, {Gloeckler},
  {Hefti}, {Judge}, {Kallenbach}, {Laeverenz}, {Lee}, {Livi}, {Managadze},
  {Marsch}, {Neugebauer}, {Ogawa}, {Reiche}, {Scholer}, {Verigin}, {Wilken}, \&
  {Wurz}}]{hilchenbach_etal:98a}
{Hilchenbach}, M., {Hsieh}, K.~C., {Hovestadt}, D., {et~al.} 1998, \apj, 503,
  916

\bibitem[{{Hilchenbach} {et~al.}(2012){Hilchenbach}, {Kallenbach}, {Hsieh}, \&
  {Czechowski}}]{hilchenbach_etal:12a}
{Hilchenbach}, M., {Kallenbach}, R., {Hsieh}, K.~C., \& {Czechowski}, A. 2012,
  in American Institute of Physics Conference Series, Vol. 1436, American
  Institute of Physics Conference Series, ed. J.~{Heerikhuisen}, G.~{Li},
  N.~{Pogorelov}, \& G.~{Zank}, 227--232

\bibitem[{{Hsieh} {et~al.}(2010){Hsieh}, {Giacalone}, {Czechowski},
  {Hilchenbach}, {Grzedzielski}, \& {Kota}}]{hsieh_etal:10a}
{Hsieh}, K.~C., {Giacalone}, J., {Czechowski}, A., {et~al.} 2010, \apjl, 718,
  L185

\bibitem[{{Huba}(2002)}]{huba:02a}
{Huba}, J.~D. 2002, {Revised NRL (Naval Research Laboratory) plasma formulary}
  (Naval Research Lab.)

\bibitem[{{Izmodenov} \& {Alexashov}(2003)}]{izmodenov_alexashov:03a}
{Izmodenov}, V.~V. \& {Alexashov}, D.~B. 2003, Astronomy Letters, 29, 58

\bibitem[{{Jokipii}(1987)}]{jokipii:87a}
{Jokipii}, J.~R. 1987, \apj, 313, 842

\bibitem[{{Krimigis} {et~al.}(2013){Krimigis}, {Decker}, {Roelof}, {Hill},
  {Amstrong}, {Gloeckler}, {Hamilton}, \& {Lanzerotti}}]{krimigis_etal:13a}
{Krimigis}, S.~M., {Decker}, R.~B., {Roelof}, E.~C., {et~al.} 2013, Science,
  341, 144

\bibitem[{{McComas} {et~al.}(2009){McComas}, {Allegrini}, {Bochsler},
  {Bzowski}, {Christian}, {Crew}, {DeMajistre}, {Fahr}, {Fichtner}, {Frisch},
  {Funsten}, {Fuselier}, {Gloeckler}, {Gruntman}, {Heerikhuisen}, {Izmodenov},
  {Janzen}, {Knappenberger}, {Krimigis}, {Kucharek}, {Lee}, {Livadiotis},
  {Livi}, {MacDowall}, {Mitchell}, {M{\"o}bius}, {Moore}, {Pogorelov},
  {Reisenfeld}, {Roelof}, {Saul}, {Schwadron}, {Valek}, {Vanderspek}, {Wurz},
  \& {Zank}}]{mccomas_etal:09a}
{McComas}, D.~J., {Allegrini}, F., {Bochsler}, P., {et~al.} 2009, Science, 326,
  959

\bibitem[{{McComas} {et~al.}(2011){McComas}, {Funsten}, {Fuselier}, {Lewis},
  {M{\"o}bius}, \& {Schwadron}}]{mccomas_etal:11a}
{McComas}, D.~J., {Funsten}, H.~O., {Fuselier}, S.~A., {et~al.} 2011, \grl, 38,
  18101

\bibitem[{{Parker}(1961)}]{parker:61a}
{Parker}, E.~N. 1961, \apj, 134, 20

\bibitem[{{Potgieter}(2013)}]{potgieter:13a}
{Potgieter}, M. 2013, Living Reviews in Solar Physics, 10, 3

\bibitem[{{Richardson} {et~al.}(2008){Richardson}, {Kasper}, {Wang}, {Belcher},
  \& {Lazarus}}]{richardson_etal:08a}
{Richardson}, J.~D., {Kasper}, J.~C., {Wang}, C., {Belcher}, J.~W., \&
  {Lazarus}, A.~J. 2008, \nat, 454, 63

\bibitem[{{Richardson} \& {Wang}(2011)}]{richardson_wang:11a}
{Richardson}, J.~D. \& {Wang}, C. 2011, \apjl, 734, L21

\bibitem[{{Rucinski} {et~al.}(1998){Rucinski}, {Bzowski}, \&
  {Fahr}}]{rucinski_etal:98a}
{Rucinski}, D., {Bzowski}, M., \& {Fahr}, H.-J. 1998, \aap, 334, 337

\bibitem[{{Stone} {et~al.}(2008{\natexlab{a}}){Stone}, {Cummings}, {McDonald},
  \& {et al.}}]{stone_etal:08a}
{Stone}, E.~C., {Cummings}, A.~C., {McDonald}, F.~B., \& {et al.}
  2008{\natexlab{a}}, in International Cosmic Ray Conference, Vol.~1,
  International Cosmic Ray Conference, 831--834

\bibitem[{{Stone} {et~al.}(2008{\natexlab{b}}){Stone}, {Cummings}, {McDonald},
  {Heikkila}, {Lal}, \& {Webber}}]{stone_etal:08b}
{Stone}, E.~C., {Cummings}, A.~C., {McDonald}, F.~B., {et~al.}
  2008{\natexlab{b}}, \nat, 454, 71

\bibitem[{{Stone} {et~al.}(2013){Stone}, {Cummings}, {McDonald}, {Heikkila},
  {Lal}, \& {Webber}}]{stone_etal:13a}
{Stone}, E.~C., {Cummings}, A.~C., {McDonald}, F.~B., {et~al.} 2013, Science,
  341, 150

\bibitem[{{Suess} \& {Nerney}(1990)}]{suess_nerney:90a}
{Suess}, S.~T. \& {Nerney}, S. 1990, \jgr, 95, 6403

\bibitem[{{Webber} \& {McDonald}(2013)}]{webber_mcdonald:13a}
{Webber}, W.~R. \& {McDonald}, F.~B. 2013, \grl, 40, 1665

\bibitem[{{Zank} {et~al.}(2013){Zank}, {Heerikhuisen}, {Wood}, {Pogorelov},
  {Zirnstein}, \& {McComas}}]{zank_etal:13a}
{Zank}, G.~P., {Heerikhuisen}, J., {Wood}, B.~E., {et~al.} 2013, \apj, 763, 20

\end{thebibliography}

\end{document}